\documentclass[amsmath,twocolumn,showpacs]{revtex4}   
\usepackage{graphicx}
\usepackage{bm}

\begin{document}

\title{Hindered mobility of a particle near a soft interface}
\author{Thomas Bickel}
\email{th.bickel@cpmoh.u-bordeaux1.fr}
\affiliation{CPMOH,
 Universit\'{e} Bordeaux 1 \& CNRS (UMR 5798) \\
351 cours de la Lib\'eration, 33405 Talence, France}

\begin{abstract}

The translational motion of a solid sphere near a deformable fluid interface
is studied in the low Reynolds number regime. 
In this problem, the fluid flow driven by the sphere is 
dynamically coupled the instantaneous conformation of the interface. 
Using a two-dimensional Fourier transform technique,
we are able to account for the multiple backflows scattered from the interface.
The mobility tensor is then obtained from the matrix elements of the relevant Green function.
This analysis allows us to express the explicit position and frequency
dependence of the mobility.
We recover in the steady limit the result for a sphere near
a perfectly flat interface. 
At intermediate time scales, the mobility exhibits an imaginary part,
which is a signature of the elastic response of the interface.
In the short time limit, we find the intriguing feature that
the perpendicular mobility may, under some circumstances,
become lower than the bulk value.
All those results can be explained from the definition
of the relaxation time of the soft interface.

\end{abstract}


\maketitle

\section{Introduction}

The motion of a particle in the vicinity of a bounding surface is 
a long standing problem in colloidal science~\cite{happelbook}.
When a colloidal sphere suspended in a quiescent fluid 
approaches a wall, the drag force acting on it
increases with respect to the drag force when far from the wall.
This property is attributed to hydrodynamic interactions that develop 
because of the boundary conditions imposed 
by the wall on the fluid flow. In addition, the motion of the particle becomes
anisotropic since the mobility is higher
in the direction parallel to the wall than in the perpendicular direction.

Although the first investigations on the 
influence of a bounding wall date back to the early work
of Lorentz~\cite{lorentz}, this field has known a certain revival during the past 
two decades. The main reason for this is certainly the achievement of
technical progress, in particular in the field of single-molecule techniques,
that allows nowadays to measure the position-dependent mobility of 
individual micron-size particles with a great accuracy. Among the most efficient tools,
one can quote evanescent waves techniques~\cite{lanPRL86,holmqvistPRE06}, 
single particle tracking by video-microscopy~\cite{faucheuxPRE94}, 
particle handling with optical tweezers~\cite{pralleAPA98,dufresnePRL00,
linPRE00}, 
AFM noise analysis~\cite{benmounaEPJE02}, 
or fluorescent correlation spectroscopy~\cite{jolyPRL06}. 
Those various methods share the common feature of
probing the random motion of
Brownian objects near one or two solid wall, and the
the mobility coefficients deduced from the experimental data agree
remarkably well with the theoretical predictions~\cite{happelbook}.

The renewal of interest for this question is also due to the 
development of microfluidics~\cite{ajdariScience05}.
Indeed, by reducing the size of the systems, the influence of surface effects
are inevitably enhanced with respect to the bulk properties. Consequently,
most of the physical phenomena take place near the boundaries. 
A fundamental understanding on how surface properties might affect the 
overall flow field has therefore become crucial in order to propose new solutions
that would take advantage of this predominance.
Lastly, it has been  suggested recently to use colloidal particles as
local probes of the flow properties near surfaces. 
This idea has been introduced in the context of the no-slip 
boundary condition~\cite{netoRPP05}, 
where the motion of the particles is expected to contain a signature 
of the slip length~\cite{jolyPRL06,laugaPF05}. More generally, one can think of a Brownian particle
as a probe of the viscoelastic properties of the bounding surface.

From a theoretical viewpoint,
the motion of a solid particle in the 
presence of a nearby, plane interface has been extensively studied
in the past. During the last few years,
calculations of mobility coefficients 
have been extended to particles near 
surfactant-covered interfaces~\cite{blawzPF99},
in a liquid film between two fluids~\cite{felderhofJCPa06},
or in a Poisseuile flow between planar walls~\cite{jonesJCP04}.
The effect of fluid inertia has also
been accounted for~\cite{felderhofJPCB05}, as well as the possibility
of liquid slippage at the wall~\cite{laugaPF05}. 
Here, we re-examine this question for a particle
near a fluid-fluid interface. Results are available
for the drag force acting on a sphere of radius
$a$ moving at a distance $z_0$ of a \textit{perfectly flat} interface, up to second
order in the ratio $a/z_0$~\cite{leeJFM79}. 
While this problem is of some intrinsic interest, and is
a logical starting point in the limit of very high surface tension,
it is obvious that a real interface will generally deform
owing to the motion of the particle.
For \textit{finite} surface tension, the motion of the 
particle is expected to be \textit{dynamically coupled}
to the conformations of the interface.
Indeed, the fluid flow caused by the 
displacement of the particle exerts stresses
that deform the interface. Relaxing back to its equilibrium
position, the interface creates a backflow that in turn
perturbs the motion of the particle, and so on.
The delay in the response of the soft surface to hydrodynamic stresses
is therefore expected to induce \textit{memory effects} in the motion
of the particle~\cite{bickelEPJE06}.

In general, the problem of the motion near a soft surface is highly non-linear due to the fact 
that the shape of the interface is unknown. 
Althought it cannot be solved exactly,
iterative solutions have been derived when the deformation of the interface 
is asymptotically small~\cite{lealJCISb82,leal1991}.
The idea is to first solve the motion of a spherical bead near a flat surface. 
As the resulting velocity produces
an imbalance of normal stress at the interface,
it is then possible to determine a first nonzero approximation for 
the deformation~\cite{lealJCISb82}.
This strategy is however limited as it only describes the 
first ``image'' correction to hydrodynamic interactions.
Also, it assumes a quasi-steady deformation profile and does not allow
for a possible delay inherent in the response of an elastic interface.

In this article, we present an analytical method that rigorously
accounts for the infinite series of hydrodynamic reflections
on the soft interface. This scheme is achieved 
within the only assumption that interface deformations remain moderate.
The rest of the paper is organized as follows. In Section~\ref{formulation}, we specify
the system and introduce the general set of equations that govern
the problem. We reformulate in Section~\ref{boundaries}
the small deformation problem in terms of equivalent boundary conditions
at the undisplaced interface. Results for the Oseen tensor and the mobility
coefficients are then discussed in Section~\ref{mobility}. In particular, we find
that the frequency-dependent mobility switches between two regimes
over a time scale corresponding to the relaxation time of the interface.
Finally, we come back to the relationship with experiments
and draw some concluding remarks in Section~\ref{discussion}.

\section{Formulation of the problem}
\label{formulation}

\subsection{Linear hydrodynamics}

We consider a spherical particle of radius $a$ moving near a fluid interface
in the low Reynolds number regime.
The interface separates two viscous, incompressible
and immiscible fluids. Its average position
is chosen to coincide with the
$x$~--~$y$ plane, with the $z$-coordinate directed perpendicular
to the it.  
The two fluids are labelled with indices
$1$ and $2$, fluid $1$ lying above fluid $2$.
Furthermore, we denote $\eta_1$ and $\eta_2$ the shear viscosities, 
$\rho_1$ and $\rho_2$ the mass densities,
and $\Delta \rho = \rho_2 -\rho_1>0$ the mass density difference.
In order to get the mobility tensor of the particle, 
we shall first evaluate the appropriate 
Green function --- called the Oseen tensor 
in this context ---
and investigate the effect of a time-dependent
point force $\mathbf{F}(t)$ acting 
at position $\mathbf{r}_0=(x_0,y_0,z_0)$ on the flow field~\cite{dhontbook}.
Without loss of generality, we can assume that
the sphere is fully immersed in fluid $1$.
For small-amplitude and low-frequency motion, 
the flow velocity  $\mathbf{v}(\mathbf{r},t)$
and the pressure $p(\mathbf{r},t)$
are governed by the Stokes equations
\begin{align}
&\eta_{\alpha} \nabla^2 \mathbf{v} -\bm{\nabla} p +\mathbf{F} \delta 
(\mathbf{r}-\mathbf{r}_0) =\mathbf{0} \ ,
\label{stokes} \\
& \bm{\nabla} .\mathbf{v} =0  \ ,
\label{inc}
\end{align}
with $\alpha=1$ or $2$, depending on whether the point $\mathbf{r}$ is 
located above or below the interface.
In Eq.~(\ref{stokes}), $\delta$ stands for the Dirac delta-distribution.
The two fluids are assumed to be quiescent except for the disturbance 
flow caused by the motion of the sphere.

\subsection{Physics of interfaces}

The Stokes equations have to be solved together with the usual 
boundary conditions at the interface,
namely the velocity and the tangential
constraints must be continuous.
The normal-normal component
of the stress tensor presents a discontinuity which is balanced
by the restoring force
exerted by the deformed interface on the fluid. 
This question is quite involved since,
in principle, the tangential and normal directions depend
on the \textit{local} and \textit{instantaneous} conformation of the interface.
However, an approximate solution can be found for moderate deformations.
In this case, the position of the 
almost flat interface can be described
by a single-valued function $h(\bm{\rho},t)$, with $\bm{\rho}=(x,y)$. 
For our purpose, it is more convenient to
use the two-dimensional Fourier representation
\begin{equation}
h(\mathbf{q},t)= \int d^2 \bm{\rho} \exp [ -i\mathbf{q}.\bm{\rho} ]
h(\bm{\rho},t) \ ,
\label{fourierspace}
\end{equation}
with $\mathbf{q}=(q_x,q_y)$. 
The elastic properties of the interface 
are then described by the Hamiltonian~\cite{safranbook}
\begin{equation}
\mathcal{H}= \frac{\gamma}{2} \int d^2 \mathbf{q} \left(
q^2 +l_c^{-2}\right) \vert h(\mathbf{q},t) \vert^2   \ ,
\label{hamiltonian}
\end{equation}
where $\gamma$ is the surface tension and 
$l_c = \sqrt{\gamma /(g\Delta \rho )} $
the capillary length,
$g$ being the gravitational acceleration.
The capillary length scale typically lies in the millimeter range
for $\gamma \approx 100$~mN/m,
but can be as low as a few microns for ultra-soft interfaces
with $\gamma \approx 0.1$~$\mu$N/m~\cite{aartsScience04}.
We then proceed in the same manner as for linearized theory of capillary waves 
and express all the boundary conditions at the undisplaced 
interface $z=0$. This hypothesis of smooth deformation is valid
up to linear order in the deformation field $h$,
so that our approach is fully  consistent with the harmonic description
of the interface energy Eq.~(\ref{hamiltonian}).

\subsection{Method of solution}

In spite of these classical simplifications, the coupling between 
the motion of the particle and the capillary waves leads to a rich behaviour.
Before solving the Stokes equations, 
we first remark that the shape of the interface depends on  
the detailed history of the motion of the particle
as well as on the shape at some earlier times.
We are then naturally lead to perform a Fourier mode
analysis in time, the Fourier transform
$\widetilde{f}(\omega)$ of an arbitrary function $f(t)$
being defined as
\begin{equation}
\widetilde{f}(\omega)= \int_{-\infty}^{+\infty} dt \exp[-i\omega t] f(t) \ .
\label{fouriertime}
\end{equation}
Besides, one can note that the 
problem is translationally invariant
along the direction parallel to the surface.
It is thus helpful to use the 
two-dimensional Fourier representation introduced above in Eq.~(\ref{fourierspace}).
It also appears judicious for this study to decompose
the vector fields into their
longitudinal, transverse and normal components~\cite{seifertAP97}.
This defines a new set of orthogonal unit vectors
$(\widehat{\mathbf{q}},\widehat{\mathbf{t}},\widehat{\mathbf{n}})$,
where $\widehat{\mathbf{q}}$ is the unit vector parallel to $\mathbf{q}$,
$\widehat{\mathbf{n}}$ the unit vector in the $z$-direction,
and $\widehat{\mathbf{t}}$ the in-plane vector perpendicular to
$\widehat{\mathbf{q}}$ and $\widehat{\mathbf{n}}$.
These vectors are connected to the cartesian basis
$(\mathbf{e}_x,\mathbf{e}_y,\mathbf{e}_z)$ through
\begin{align}
\widehat{\mathbf{q}} &=\frac{q_x}{q}\mathbf{e}_x
+  \frac{q_y}{q}\mathbf{e}_y  \ , \nonumber \\
\widehat{\mathbf{t}} &=\frac{q_y}{q}\mathbf{e}_x
-  \frac{q_x}{q}\mathbf{e}_y   \ , \\
\widehat{\mathbf{n}} &=\mathbf{e}_z  \ . \nonumber
\end{align}
The velocity and the force are written 
$ \mathbf{v} = v_l \widehat{\mathbf{q}}
 +v_t\widehat{\mathbf{t}}
+v_z\widehat{\mathbf{n}}$ and $\mathbf{F}=
F_l \widehat{\mathbf{q}} 
+F_t\widehat{\mathbf{t}} + F_z\widehat{\mathbf{n}}$, respectively.
Inserting these representations into the Stokes 
equations~(\ref{stokes})~--~(\ref{inc})
finally leads to a  system of ordinary differential equations
for the Fourier-transformed quantities
\begin{eqnarray}
-\eta_{\alpha} q^2 \widetilde{v}_l  + \eta_{\alpha} \displaystyle{\frac{\partial^2 \widetilde{v}_l }{\partial z^2}} 
& -  i q \widetilde{p}  &+ \widetilde{F}_l \delta (z-z_0)  =  0   \label{stokesl} \\
-\eta_{\alpha} q^2 \widetilde{v}_t + \eta_{\alpha} \displaystyle{\frac{\partial^2 \widetilde{v}_t }{\partial z^2}} 
&&+ \widetilde{F}_t \delta (z-z_0) = 0   \label{stokest}    \\
-\eta_{\alpha} q^2 \widetilde{v}_z +\eta_{\alpha} \displaystyle{\frac{\partial^2 \widetilde{v}_z }{\partial z^2}}
&-\displaystyle{\frac{\partial \widetilde{p}}{\partial z}}& +\widetilde{F}_z \delta(z-z_0) =0  \label{stokesn}
\end{eqnarray}
with the divergenceless condition
\begin{equation}
i q \widetilde{v}_l +\frac{\partial \widetilde{v}_z}{\partial z} =0 \ .
\label{incompressibility}
\end{equation}
Although this framework is not as ``transparent'' as the usual
image method, its advantages are twofold. On the one hand,
it is particularly 
well suited to accommodate with the description of 
the interface energy in Fourier space, since it thoroughly accounts for
the symmetries of the problem.
On the other hand, 
the transverse component of the velocity is  decoupled from
the longitudinal and normal directions.
Moreover, relation~(\ref{incompressibility})  provides
a usefull link between $\widetilde{v}_l$ and
$\widetilde{v}_z$, so that it is not difficult to get a single,
fourth order differential equation for
the normal component only 
\begin{equation}
\frac{\partial^4 \widetilde{v}_z }{\partial z^4} - 2q^2
\frac{\partial^2 \widetilde{v}_z }{\partial z^2}
+q^4  \widetilde{v}_z
 =  \frac{q^2\widetilde{F}_z}{\eta_1 } \delta (z-z_0) 
+\frac{i q \widetilde{F}_l}{\eta_1 } \delta' (z-z_0) \ .
\label{eqtextvz}
\end{equation}
Here, $\delta'$ is the derivative of the Dirac delta-distribution.

\section{Boundary conditions}
\label{boundaries}

To describe the flow in the presence of an interface, we must
consider the flow on each side separately,
and then require proper matching conditions for the velocity
and surface forces.
As stated above, the hypothesis of smooth deformations around
the planar configuration enables us
to express the boundary conditions at the undisplaced interface $z=0$~\cite{lealJCISb82}. 
Because our representation of the velocity in terms of longitudinal
and transverse coordinates is not commonly used in the litterature, 
we find it worthwhile to give some details regarding the derivation
of the boundary values.

\subsection{Continuity of the velocity}

First of all, we have to ensure that the velocity is 
continuous at the interface. Explicitly,
this requirement reads
\begin{eqnarray}
\widetilde{v}_l (\mathbf{q},0^+,\omega) = 
\widetilde{v}_l (\mathbf{q},0^-,\omega) \ , \label{vl0} \\
\widetilde{v}_t (\mathbf{q},0^+,\omega) = 
\widetilde{v}_t (\mathbf{q},0^-,\omega) \ , \label{vt0} \\
\widetilde{v}_z (\mathbf{q},0^+,\omega) = 
\widetilde{v}_z (\mathbf{q},0^-,\omega)  \ . \label{vz0}
\end{eqnarray}
Interestingly, the condition~(\ref{vl0}) for the \textit{longitudinal}
coordinate together with the 
incompressibility condition~(\ref{incompressibility}) implies
an additional boundary condition for the \textit{normal} coordinate
of the velocity, namely
\begin{equation}
\frac{\partial \widetilde{v}_z}{\partial z} \Big\vert_{0^+} =
 \frac{\partial \widetilde{v}_z}{\partial z} \Big\vert_{0^-}  \ .
\label{dvz0}
\end{equation}

\subsection{Balance of tangential forces}

Secondly, tangential stresses have to be balanced at the interface.
In real space, the continuity condition for the normal-tangential 
components of the stress tensor reads $\sigma_{zx} \vert_{0^+}
= \sigma_{zx} \vert_{0^-}$ and $\sigma_{zy} \vert_{0^+}
= \sigma_{zy} \vert_{0^-}$, with 
$\sigma_{jk}=-p \delta_{jk} + \eta_{\alpha}
(\partial v_j / \partial x_k +
\partial v_k / \partial x_j)$ the stress tensor
in cartesian coordinates. Switching to $\{\mathbf{q},
z,\omega \}$ variables, both requirements reduce to
\begin{equation*}
\eta_1 \left( \frac{\partial \widetilde{\mathbf{v}}_{\parallel}}{\partial z}
+i\mathbf{q} \widetilde{v}_z \right) \Big\vert_{0^+}
=\eta_2 \left( \frac{\partial \widetilde{\mathbf{v}}_{\parallel}}{\partial z}
+i\mathbf{q} \widetilde{v}_z \right) \Big\vert_{0^-}  \ ,
\end{equation*}
where the two-dimensional vector
$\widetilde{\mathbf{v}}_{\parallel} = (\widetilde{v}_l ,
\widetilde{v}_t )$ is the parallel velocity. 
Projecting this equation onto the transverse
direction leads to the condition
\begin{equation}
\eta_1  \frac{\partial \widetilde{v}_t}{\partial z}
 \Big\vert_{0^+}
=\eta_2  \frac{\partial \widetilde{v}_t}{\partial z}
 \Big\vert_{0^-}  \ ,
\label{dvt0}
\end{equation}
whereas  projection onto the longitudinal coordinate gives
another condition which still involves both $\widetilde{v}_l$
and $\widetilde{v}_z$.
In order to obtain a boundary condition for the normal component only, 
the incompressibility condition~(\ref{incompressibility}) is once more
invoked. We finally get
\begin{equation}
\eta_1 \left(  \frac{\partial^2 \widetilde{v}_z}{\partial z^2}
+q^2 \widetilde{v}_z \right) \Big\vert_{0^+}
=\eta_2 \left( \frac{\partial^2 \widetilde{v}_z}{\partial z^2}
+q^2 \widetilde{v}_z \right) \Big\vert_{0^-}  \ .
\label{ddvz0}
\end{equation}
Note that the balance of \textit{tangential} stresses is also relevant
with regard to the \textit{normal} component of the velocity.

\subsection{Discontinuity of normal stress}

The next condition that has to be enforced concerns the normal-normal stress difference
that comes into play whenever the interface is bent. 
Indeed, a deformation of the interface gives rise to normal
restoring forces, expressed as the functional derivative of the 
Hamiltonian~(\ref{hamiltonian}).
For small displacements, the forces are small and 
proportionnal to $h$. 
The normal stress condition reads, in real space,
$\sigma_{zz} \vert_{0^-} - \sigma_{zz} \vert_{0^+}
= - \delta \mathcal{H}/\delta h$. In terms of the variables
$\{\mathbf{q},z,\omega\}$, we have
\begin{equation*}
 \widetilde{p}(0^+)-\widetilde{p}(0^-)   - 
2\eta_1  \frac{\partial \widetilde{v}_z}{\partial z}
  \Big\vert_{0^+}
+2\eta_2  \frac{\partial \widetilde{v}_z}{\partial z}
  \Big\vert_{0^-}
=-E_q \widetilde{h}(\mathbf{q},\omega)  \ ,
\end{equation*}
where we define the energy density $E_q=\gamma (q^2+\lambda^{-2})$.
It can be seen that the normal stress difference at the interface
is balanced by interfacial tension and  buoyancy forces
(due to the density difference between the two fluids).
This condition still involves both the normal component of the velocity
as well as the pressure field. To get a relation in terms of 
$\widetilde{v}_z$ only, we shall first use Eq.~(\ref{stokesl})
to express the pressure difference  
(remember that $z_0 >0$) 
\begin{align*}
iq & \left( \widetilde{p}(0^+)-  \widetilde{p}(0^-) \right)    \\
&= \eta_1 \left( \frac{\partial^2 \widetilde{v}_l}{\partial z^2}
-q^2\widetilde{v}_l \right)  \Big\vert_{0^+}  
 -\eta_2 \left( \frac{\partial^2 \widetilde{v}_l}{\partial z^2}
-q^2\widetilde{v}_l \right)  \Big\vert_{0^-}  \ .
\end{align*}
Substitute $\widetilde{v}_l$ for
$\widetilde{v}_z$ thanks to relation~(\ref{incompressibility}), 
we arrive at the condition
on the third derivative of the velocity
\begin{align}
  \eta_1  &\left(  \frac{\partial^3 \widetilde{v}_z}{\partial z^3} 
- 3q^2   \frac{\partial \widetilde{v}_z}{\partial z}
\right)\Big\vert_{0^+}   \nonumber \\
& = \eta_2 \left(
   \frac{\partial^3 \widetilde{v}_z}{\partial z^3}  
 -3q^2   \frac{\partial \widetilde{v}_z}{\partial z} 
\right)\Big\vert_{0^-}  
 -q^2 E_q \widetilde{h}(\mathbf{q},\omega)  \ .
\label{dddvz0}
\end{align}

\subsection{Immiscibility of the two fluids}

To make the calculations tractable, we suppose that 
the condition of immiscibility can be written at $z=0$.
This approximation is justified since the fact that it 
is in any rigour valid at $z=h$ is an effect of higher order.
Within this assumption, the time rate of change
of the shape function is related to the normal velocity at the interface through
\begin{equation}
\widetilde{v}_z (\mathbf{q},0,\omega) = 
i \omega \widetilde{h}(\mathbf{q},\omega)  \ ,
\label{closure}
\end{equation}
up to linear order in the deformation field.
This closure relation is especially relevant since,
as shown in the following, it allows to work out
the instantaneous deformation of the interface
in response to hydrodynamic stresses.

\section{Green function and translational mobility}
\label{mobility}

\subsection{Motion of the interface}

We now have all the ingredients to solve the Stokes equations.
Because our calculations are algebraically involved, 
we save the details for the appendices. 
An interesting results is that the local deformation
of the interface is directly proportional to the amplitude of the point
force applied at height $z_0$
\begin{equation}
\widetilde{h}(\mathbf{q},\omega) = \mathbf{\widetilde{R}}(\mathbf{q},z_0,\omega)
  \widetilde{\mathbf{F}}(\omega)     \ ,
\label{response}
\end{equation}
where the vector $\mathbf{\widetilde{R}}$ is the response function obtained 
thanks to the closure relation~(\ref{closure}). 
For a vertical force 
$\widetilde{\mathbf{F}}= (0,0,\widetilde{F}_z)$, 
we find in Appendix~\ref{zj}
\begin{equation}
\widetilde{R}_z(\mathbf{q},z_0,\omega) = \frac{1}{4  \overline{\eta} q \left(\omega_q + 
i \omega\right)} (1+qz_0) e^{-qz_0}    \ .
\label{defz}
\end{equation}
As expected, the relaxation dynamics of the 
profile is governed by the \textit{mean} viscosity 
$\overline{\eta}= (\eta_1 + \eta_2)/2$. 
The response of a deformation mode with wavevector $\mathbf{q}$ 
is characterized by its relaxation rate 
\begin{equation}
\omega_q = \frac{\gamma}{4q \overline{\eta}} 
\left( q^2 +l_c^{-2} \right)  \ .
\end{equation}
Remark that differents wavevectors are not damped in the same way.
The amplitude of the response function is always maximum for $\mathbf{q}=\mathbf{0}$,
$\widetilde{h}(\mathbf{0},\omega)=\widetilde{F}_z/(\Delta \rho g)$.
It then vanishes with increasing $q$, all the more rapidly as
the frequency $\omega$ or the distance $z_0$ are large.
The result Eq.~(\ref{defz}) can be interpreted as follows.
The real part of $\widetilde{R}_z$, which is in phase with
the strain, is the analogous of a storage modulus for a
viscoelastic medium~\cite{larsonbook}.
This contribution corresponds to the elastic energy stored
in the deformation of the interface. On the other hand, the imaginary part of
$\widetilde{R}_z$ plays the role of a loss modulus
and describes the viscous dissipation associated with
the relaxation of individual deformation modes. 

Coming back to the motion of the interface in real space,
one can evaluate the inverse Fourier transform of the
response function, though the calculations will not be performed here.
A deformation may also be obtained as a result of 
a point force applied parallel to the interface. We find
\begin{align}
\widetilde{R}_l(\mathbf{q},z_0,\omega)  &= \frac{1}{4  \overline{\eta} q \left(\omega 
-i \omega_q \right)} qz_0 e^{-qz_0}  \ , \\  
\widetilde{R}_t(\mathbf{q},z_0,\omega)  &=0\ ,
\label{defp}
\end{align}
for the longitudinal and transverse coordinate, respectively.
Note that the shape of the interface is
not affected by the transverse component of the force.

\subsection{Oseen tensor}

The components of the Oseen tensor 
are then obtained by identification with the definition 
\begin{equation}
\widetilde{v}_i=  \sum_j \widetilde{\mathcal{G}}_{ij}  \widetilde{F}_j  \ ,
\end{equation}
where $i,j \in \{l,t,z\}$.  For symmetry reasons, 
the Green function satisfies the general relation
$\mathcal{G}_{ij}(\mathbf{r},\mathbf{r}',t)
= \mathcal{G}_{ji}(\mathbf{r}',\mathbf{r},t)$
for $z$ and $z'$ on the same side of the interface~\cite{pozbook}.
This is re-expressed in terms of our 
particular choice of variables as
$\widetilde{\mathcal{G}}_{ij}(\mathbf{q},z,z',\omega)
= \widetilde{\mathcal{G}}_{ji}(-\mathbf{q},z',z,-\omega)^*$,
property that can be checked along the calculations.
As shown in the appendices, the Green function can always be written as
\begin{align}
\widetilde{\mathcal{G}} (\mathbf{q},z,z_0,\omega) &= 
\widetilde{\mathcal{G}}^{(0)} (\mathbf{q},z-z_0)+
\Delta \widetilde{\mathcal{G}} (\mathbf{q},z,z_0,\omega) \nonumber \\
&= \widetilde{\mathcal{G}}^{(0)} (\mathbf{q},z-z_0)+
\Delta \widetilde{\mathcal{G}}^{(1)} (\mathbf{q},z,z_0)  \nonumber \\
&+ \frac{\omega}{\omega -i\omega_q}
\Delta \widetilde{\mathcal{G}}^{(2)} (\mathbf{q},z,z_0)  \ ,
\label{decg}
\end{align}
see Appendix~\ref{tj} for the exact expression of 
$\widetilde{\mathcal{G}}_{tj}$,
Appendix~\ref{zj} for the components $\widetilde{\mathcal{G}}_{zj}$,
and Appendix~\ref{lj} for the components $\widetilde{\mathcal{G}}_{lj}$.
The first term, which
depends only the relative distance $(z-z_0)$, would 
reduce to the usual free-space
Green function if the viscosities were  equal.
The second term, $\Delta \widetilde{\mathcal{G}}^{(1)}$,
is the correction for an undistorted interface. 
Both contributions have already been obtained
in previous work, though not in this particular choice of coordinates~\cite{leeJFM79}.
The original part of this study is the derivation the contribution
coming from the \textit{deformation} of the interface. 
The prefactor $(\omega -i\omega_q)^{-1}$ in Eq.~(\ref{decg})
is a clear signature of hydrodynamic scattering effects on the soft surface.
Indeed, the fluid flow resulting from a displacement of the 
particle exerts stresses that deform the interface. Relaxing back to its
equilibrium position, the interface creates a backflow that in 
turn perturbs the motion of the particle, and so forth. 
The infinite sum $(\omega -i\omega_q)^{-1}=
i/\omega_q\sum_{n=0}^{\infty} (-i)^n(\omega / \omega_q)^n$
is the expression of this infinite series of ``reflexions'' of the 
original point force on the interface.
This argument is confirmed by the fact that
the correction vanishes 
for $\gamma \rightarrow \infty$. 
One recovers the results for the flat
liquid-liquid interface
in the high surface tension limit, as expected.

Finally, once all the components are known
in the $(\widehat{\mathbf{q}},\widehat{\mathbf{t}} ,
\widehat{\mathbf{n}})$ basis, it is not difficult to express
the Oseen tensor in cartesian coordinates. 
In particular, the diagonal components are given by
\begin{align}
\widetilde{\mathcal{G}}_{xx}  &=  \frac{q_x^2}{q^2}
\widetilde{\mathcal{G}}_{ll} +  \frac{q_y^2}{q^2}
\widetilde{\mathcal{G}}_{tt} \ , \\
\widetilde{\mathcal{G}}_{yy}  &=  \frac{q_y^2}{q^2}
\widetilde{\mathcal{G}}_{ll}  +  \frac{q_x^2}{q^2}
\widetilde{\mathcal{G}}_{tt} \ .
\end{align}
Similar relations can be deduced for off-diagonal terms,
though they will not be required in the following.

\subsection{Translational mobility tensor}

From the matrix elements of the Oseen tensor, we can obtain the mobility matrix for a sphere. 
To this aim, we still have to enforce the
no-slip boundary condition for the fluid flow on the surface of the particle. 
In the following, we assume that the particle is a sphere of radius $a$.
If we note $\mathbf{U}(\mathbf{r}_0)$
and $\boldsymbol{\Omega}$ respectively the translational and rotational velocity 
of the sphere, $\mathbf{r}_0$ being the position of its center-of-mass,
then the fluid velocity satisfies 
\begin{equation}
\mathbf{v}(\mathbf{r}_0+\mathbf{a}) = \mathbf{U}(\mathbf{r}_0) + 
\boldsymbol{\Omega} \times \mathbf{a}   \ ,
\label{clhydro}
\end{equation}
for any vector $\mathbf{a}$ scanning the surface of the bead.
Integrating this relation over the particle,
one obtains a linear relation
between the total friction force $ \mathbf{F}_H $ exerted by the liquid 
and velocity of the particle~\cite{dhontbook}.
This relation defines the (frequency-dependent) mobility tensor through
$\widetilde{\mathbf{U}} = - \widetilde{\mu} 
\widetilde{ \mathbf{F}}_H  $.
It can be 
written as the sum of two terms,
$\widetilde{\mu}_{kl} (z_0,\omega) = \mu_0 \delta_{kl} + \Delta 
\widetilde{\mu}_{kl} (z_0,\omega)$, with $\mu_0=(6\pi \eta_1 a)^{-1}$
the bulk value for a particle in fluid 1 but infinitely far from the interface. 
The correction $\Delta \widetilde{\mu}_{kl} $ 
is then expanded in powers of $a/z_0$.
In the limit of small particles $a \ll z_0$,
the correction to the mobility tensor is given,
at leading order, by
\begin{equation}
\Delta \widetilde{\mu}_{kl} \left( z_0,\omega \right)=\int \frac{d^2\mathbf{q}}{(2\pi)^2}
\Delta \widetilde{\mathcal{G}}_{kl}(\mathbf{q},z_0,z_0,\omega ) \ .
\label{deltamu}
\end{equation}
As a matter of fact, it can be shown that all cross-contributions
vanish, so that the correction to the mobility tensor is also diagonal with
elements $ \Delta \widetilde{\mu}_{xx} =  \Delta \widetilde{\mu}_{yy}
= \Delta \widetilde{\mu}_{\parallel}$ and 
$\Delta \widetilde{\mu}_{zz}= \Delta \widetilde{\mu}_{\perp}$.

\subsubsection{Perpendicular mobility}

\begin{figure}
\centering
\includegraphics[width=3.25in]{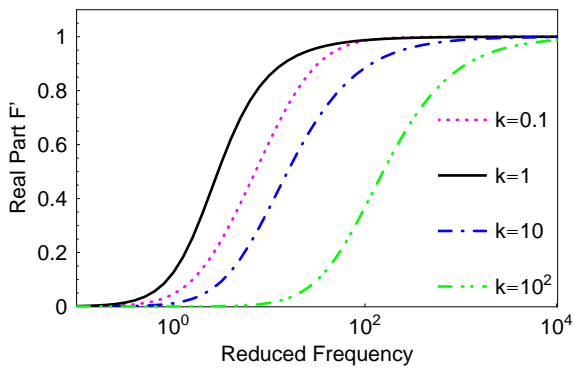}
\caption{(Color online) Real part of $F$ as a function of the reduced frequency $s=\omega \tau$,
for different values of the reduced distance $k=z_0/l_c$. The position of the cross-over between
the two regimes $\omega \tau \ll 1$ and $\omega \tau \gg 1$ is quite sensitive to the distance
to the interface.}
\label{fig1}
\end{figure}

\begin{figure}
\centering
\includegraphics[width=3.25in]{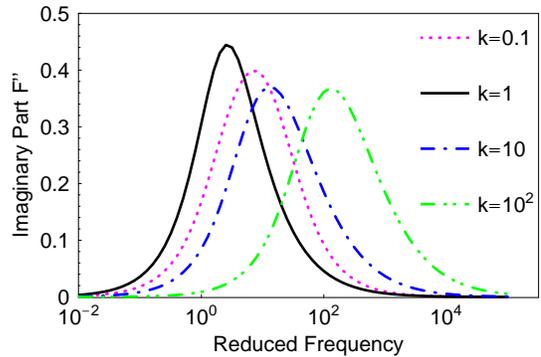}
\caption{(Color online) Imaginary part of $F$ as a function of the reduced frequency $s=\omega \tau$,
for different values of the reduced distance $k=z_0/l_c$. The elastic coupling is maximum
around the value $\omega \tau \sim 1$.}
\label{fig2}
\end{figure}

From the result Eq.~(\ref{gzzsup}) for the normal-normal component of the Oseen
tensor, we find 
\begin{align}
\Delta \widetilde{\mu}_{\perp} \left( z_0,\omega \right) =
 -  \frac{1}{16 \pi \eta_1 z_0}    &  \left(\frac{2\eta_1+3\eta_2}{\eta_1 +\eta_2}\right) \nonumber \\
 &+  \frac{5}{ 32\pi \overline{\eta} z_0 } 
F  \left( \omega \tau , \frac{z_0}{l_c} \right) \ .
\label{deltamuzz}
\end{align}
In this expression, $\tau= 4 \overline{\eta} l_c/\gamma$ 
corresponds to the longest time 
required for elastic stuctures
in the fluid --- in our case, the interface ---
to relax. For typical values $\overline{\eta} = 10^{-2}$~Pa.s
and $\Delta \rho = 10^2$~kg.m$^{-3}$,
it ranges from $\tau \approx 10^{-3} $~s for usual interfaces
with $\gamma=100$~mN.m$^{-1}$
up to $\tau \approx 1$~s  for ultra-soft interfaces
with $\gamma=0.1$~$\mu$N.m$^{-1}$~\cite{aartsScience04}.
The frequency-dependent contribution $F$ arises from surface deformations
and is therefore governed by the mean viscosity $\overline{\eta}$.
It is given by
\begin{align}
F(s,k) &= \frac{4}{5}\int_0^{\infty} dx \frac{iksx}{1+isx+x^2}
(1+k x)^2 \exp[-2k x] \nonumber \\
&= F'(s,k) +i F''(s,k)  \ .
\label{formula}
\end{align}
This integral actually corresponds to the sum over all 
deformation modes of the interface~\cite{remark}.
For $\omega =0$, one has
$F(0, z_0/l_c)=0$ 
and the authoritative reader will 
recognize on the right hand side of Eq.~(\ref{deltamuzz}) 
the correction to the mobility
of a sphere near a flat, liquid-liquid interface~\cite{leeJFM79}.
One even recovers the result of Lorentz for a hard wall
by taking the limit $\eta_2 \rightarrow \infty$. 
For finite values of $\omega$, the additional term is
actually a complex number. Its real part $F'$ 
represents the contribution to the viscous dissipation that comes
from interface deformations. 
As shown in Fig.~\ref{fig1}, $F'$ is
positive for any value of the parameters, so that 
the real part of the mobility increases when 
the constraint on the shape of the interface is released. 
Viscous dissipation is therefore \textit{always
lower} for a soft interface, which can bend under
hydrodynamic forces, than for a rigid interface.
Another outcome is that the mobility of the particle also
exhibits an imaginary part $F''$, which corresponds to the storage of elastic
energy in the deformation of the interface. 
As shown on Fig.~\ref{fig2}, $F''$ is non-zero only
for intermediate values of the frequency $\omega \tau \sim 1$.
The latter contribution vanishes when $\omega \rightarrow \infty$ and
one gets in this limit
\begin{equation}
\Delta \widetilde{\mu}_{\perp} \left( z_0,\omega \rightarrow \infty \right)=
\frac{3}{16 \pi \eta_1 z_0}
\left(\frac{\eta_1-\eta_2}{\eta_1 +\eta_2}\right) \ .
\label{cordeltamuzz}
\end{equation}
Lastly, we remark from Fig.~\ref{fig1} and~\ref{fig2} that
both $F'$ and $F''$ are of $\mathcal{O}(1)$ for a wide range of
reduced distances $z_0/l_c$. But because of the
prefactor $z_0^{-1}$ in Eq.~(\ref{deltamuzz}),
the coupling between the motion of the particle
and the shape of the interface vanishes when the particle
is far away from the surface,
as one might expect. 

To better understand those results, let us focus on the
relaxation time $\tau$. From the definition 
$ \tau = 4 \overline{\eta} l_c / \gamma \propto 1/\sqrt{\gamma}$,
it can be noticed that the limit $\omega \tau \ll 1$
actually coincides with the limit $\gamma \rightarrow \infty$.
At very low frequencies, the interface thus appears infinitely
rigid and one therefore recovers the mobility of a particle near
a flat interface. This also explains why the imaginary part 
of the correction vanishes when $\omega=0$. 
A similar reasoning applies to the other limit $\omega \tau \gg 1$,
that would correspond to an interface
with vanishing surface tension. Since no elastic energy can be stored
anymore, $F''$ has to vansih when $\omega \rightarrow \infty$.
Interestingly, the \textit{sign} 
of the real part of $\Delta \widetilde{\mu}_{\perp}$  may change depending
on whether $\omega \tau \ll 1$ or 
$\omega \tau \gg 1$. Indeed, it is always negative
at low frequencies, whereas it may be positive at high frequencies
provided that $\eta_1>\eta_2$. This susprising property,
peculiar to soft interfaces, may strongly influence the 
statistical properties of Brownian particles since surface deformation
may enhance diffusion -- with regards to the bulk value -- at short times.

\subsubsection{Parallel mobility}

Similar conclusions can be drawn for 
the mobility parallel to the surface.
We find that
\begin{align}
\Delta \widetilde{\mu}_{\parallel} \left( z_0,\omega \right)=
\frac{1}{32 \pi \eta_1 z_0}  & \left(\frac{2\eta_1-3\eta_2}{\eta_1 +\eta_2}\right) \nonumber \\
& +\frac{1}{ 64\pi \overline{\eta} z_0 } 
G  \left( \omega \tau , \frac{z_0}{l_c} \right) \ ,
\label{deltamuxx}
\end{align}
where the frequency-dependent contribution is given by
\begin{equation}
G(s,k) = 4\int_0^{\infty} dx \frac{ iksx }{1+isx+x^2}
k^2 x^2  \exp[-2k x] \ .
\label{formulax}
\end{equation}
In particular, one recovers the mobility coefficient for
sphere near a rigid interface in the asymptotic limit
$\omega \tau \ll 1$ 
\begin{equation}
\Delta \widetilde{\mu}_{\parallel} \left( z_0,\omega =0 \right)=
\frac{1}{32 \pi \eta_1 z_0}
\left(\frac{2\eta_1-3\eta_2}{\eta_1 +\eta_2}\right) \ ,
\label{limmuxx}
\end{equation}
whereas one obtains in the other limit $\omega \tau \gg 1$
\begin{equation}
\Delta \widetilde{\mu}_{\parallel} \left( z_0,\omega \rightarrow \infty \right)=
\frac{3}{32 \pi \eta_1 z_0}
\left(\frac{\eta_1-\eta_2}{\eta_1 +\eta_2}\right) \ .
\label{cordeltamuxx}
\end{equation}

\section{Discussion}
\label{discussion}

To summarize, we have calculated the mobility
tensor of a spherical particle moving
close to a fluid-fluid interface.
Several lengths are
inherent in the system, namely the radius $a$ of the 
particule, the distance from the wall $z_0$, and the 
capillary length $l_c$. The results presented in this work concern
the response to a point force and are valid for
particles far from the interface $a \ll z_0$.
Because a soft interface can deform and store elastic energy,
the mobility tensor decomposes into a real 
and an imaginary part. 
In steady-state limit $\omega \tau \ll 1$, deformations are irrelevant and
one recovers the classical result 
for a flat, fluid-fluid interface. 
On the other hand, the short-time limit $\omega \tau \gg 1$ 
presents the intriguing feature that the perpendicular mobility
can be higher than the bulk mobility if $\eta_1>\eta_2$. Yet this result does not break
any fundamental law since it arises from the fact that the particle
``feels'' the other side of the interface, which has a lower shear viscosity.
Finally coming back to time variable, the friction force 
experienced by the particle will be expressed as a convolution product and is
therefore non-local in time.  Solvent backflow and
delay of the response of the elastic interface then induce \textit{memory effects}
in the motion of the particle.

The framework developed in this study may be adapted
to various problems near soft interfaces.
For instance, one
might investigate surface-mediated contributions to
the coupled diffusion of two particles.
One can also consider more complex surfaces, such as 
surfactant-covered interfaces or fluid membranes.
Predictions regarding the rotational mobility might  be 
relevant for experiments as well, especially in the case of anisotropic particles.
Remark that translational and rotational motions are not coupled
for a sphere in the linearized theory. This might not be true anymore
for large deformations, where non-linear effects come into play~\cite{lealJCISa82}.

Another point that might be included in the theory
is the effect of fluid inertia. This contribution has been
neglected so far, though it becomes relevant
at frequencies higher than $\omega_c = \eta /(\rho a^2)$. 
For typical values $\eta = 10^{-3}$ Pa.s, $\rho =10^3$ kg.m$^{-3}$
and $a=1$ $\mu$m, we obtain $\omega_c \approx 10^6$ rad.s$^{-1}$.
Here however, we consider time scales 
comparable to the relaxation time of the interface. 
This corresponds to frequencies in the kH range, so that our 
approximation is fully justified.
At this point, it should be mentioned that a study similar to ours,
including fluid inertia, has been published during the completion
of this work~\cite{felderhofJCPb06}. In the steady limit, the author obtains
the result for a \textit{rigid wall} with stick boundary conditions.
This however cannot be correct since one expects to find in this limit 
the mobility of a sphere near a \textit{fluid-fluid} interface.
The results derived in Ref.~\cite{felderhofJCPb06} are therefore questionable,
but a closer inspection would be required 
to identify the origin of the discrepancy.

Finally, let us comment on some possible comparisons with experiments. 
Recently, de Villeneuve \textit{et al.} have considered
the sedimentation of PMMA spheres towards an interface with
ultra-low tension $\gamma \approx 0.1$~$\mu$N/m~\cite{villeneuveCSA06}.
In this regime, long-range hydrodynamic interactions are dominant
and lubrication theory does not apply.
The authors clearly observe strong deformations of the interface,
of the order several micrometers for spheres with radius 
$a=15$~$\mu$m~\cite{villeneuveCSA06}.
Moreover, they measure sedimentation velocities that
do not follow the theoretical curves for an undistorted interface,
the particles falling faster towards
the soft interface.
The interpretation of those results might be quite
straightforward in the light of the 
present analysis, even though the 
non-linear equations of motion
might be challenging to solve. 
Work on this question is currently under progress.

\acknowledgements

D. Aarts, L. Bocquet, I. Pagonabarraga, and
V. de Villeneuve are gratefully acknowledged for useful discussions.
The author also wishes to thank S. Villain-Guillot
and A. W\"urger for most valuable comments.

\appendix

\section{Transverse component of the velocity}
\label{tj}

We begin with Eq.~(\ref{stokest}) for the transverse component, which 
is easier to solve since it does not
couple with the longitudinal and vertical coordinates
of the velocity.
This equation can be rewritten as
\begin{equation}
\frac{\partial^2 \widetilde{v}_t }{\partial z^2} - q^2 \widetilde{v}_t
 = - \frac{\widetilde{F}_t}{ \eta_1 q^2  } \delta (z-z_0)   \ .
\end{equation}
With the condition that the fluid is at rest at infinity,
the solution is
\begin{align*}
&\widetilde{v}_t (\mathbf{q},z,\omega) =  A e^{-qz}   & \mbox{for } & 0<z_0<z   \ ,\\
&\widetilde{v}_t (\mathbf{q},z,\omega) =  Be^{qz} + C e^{-qz}   & \mbox{for } & 0<z<z_0 \ , \\
&\widetilde{v}_t (\mathbf{q},z,\omega) =  De^{qz}  & \mbox{for } & z<0<z_0  \ .
\end{align*}
We then need to specify the boundary conditions in order to
determine the four integration constants. 
The continuity of
the velocity and the balance of tangential stresses
at height $z=0$ give the conditions~(\ref{vt0}) and~(\ref{dvt0}).
We get another couple of conditions by invoking the
standard continuity conditions for the Green 
function at the singularity $z=z_0$. Explicitely, these requirements read
\begin{align}
& \widetilde{v}_t(\mathbf{q},z_0^+,\omega) = \widetilde{v}_t(\mathbf{q},z_0^-,\omega)  \ ,
\label{bcz1t}\\
&   \frac{\partial \widetilde{v}_t}{\partial z} \Big\vert_{z_0^+} - 
 \frac{\partial \widetilde{v}_t}{\partial z} \Big\vert_{z_0^-}
= -\frac{\widetilde{F}_t}{ \eta_1 q^2 }\ .
\label{bcz2t}
\end{align}
Enforcing the boundary conditions~(\ref{vt0}), (\ref{dvt0}),
(\ref{bcz1t}) and~(\ref{bcz2t}),
we find the following expression for $z \geq 0$
\begin{equation}
\widetilde{v}_t (\mathbf{q},z,\omega) =  \frac{\widetilde{F}_t}{2  \eta_1 q }
\left[  e^{-q\vert z-z_0\vert } - \left( \frac{1- \lambda}{1+\lambda } \right)
e^{-q(z+z_0)} \right] \ ,
\label{vtrans1}
\end{equation}
and for $z\leq 0$
\begin{equation}
\widetilde{v}_t (\mathbf{q},z,\omega) =  \frac{\widetilde{F}_t}{2  \eta_2 q}
 \left( \frac{2}{1+\lambda } \right)
e^{-q\vert z-z_0\vert}  \ .
\label{vtrans2}
\end{equation}

The transverse components of the Green function are then 
obtained by comparison of equation~(\ref{vtrans1}) (for $z \geq 0$)
or~(\ref{vtrans2}) (for $z \leq 0$)
with the definition of the Oseen tensor
$ \widetilde{v}_t=  \widetilde{\mathcal{G}}_{tl}  \widetilde{F}_l +
\widetilde{\mathcal{G}}_{tt} \widetilde{F}_t 
+ \widetilde{\mathcal{G}}_{tz}  \widetilde{F}_z $.
Obviously, we get $ \widetilde{\mathcal{G}}_{tl}
= \widetilde{\mathcal{G}}_{tz}=0$,
the  only non-zero component being $ \widetilde{\mathcal{G}}_{tt}$. 
Note that the transverse component of the velocity is
not affected by the shape of the interface.

\section{Normal component of the velocity}
\label{zj}

\subsection{Differential equation and general solution}

The solution of the fourth-order differential equation~(\ref{eqtextvz})
satisfied by $\widetilde{v}_z$  is 
\begin{align*}
\widetilde{v}_z (\mathbf{q},z,\omega) =&  \left(A  +Bz\right) e^{-qz}    &  \mbox{for } &  z>z_0 \\
\widetilde{v}_z (\mathbf{q},z,\omega) =&  \left(C  +Dz\right) e^{qz} + 
\left(E +Fz \right) e^{-qz}  & \mbox{for }  &z<z_0 \\
\widetilde{v}_z (\mathbf{q},z,\omega) =&  \left(G  +Hz\right) e^{qz}   & \mbox{for } &  z<0
\end{align*}
For the sake of simplicity, 
we shall focus separately 
on the situations where $(\widetilde{F}_l = 0,\widetilde{F}_z \neq 0)$
and  $(\widetilde{F}_l \neq 0,\widetilde{F}_z = 0)$.
According to the superposition 
principle, each solution leads by identification to the components
of the Green tensor $\widetilde{\mathcal{G}}_{zz}$ and 
$\widetilde{\mathcal{G}}_{zl}$ , respectively.
Obviously, the normal-transverse component
is identically zero, $\widetilde{\mathcal{G}}_{zt}=0$.

\subsection{Normal-normal component}

We first consider the case where $\widetilde{F}_l = 0$
and $\widetilde{F}_z \neq 0$.
In addition to the boundary conditions~(\ref{vz0}), (\ref{dvz0}),
(\ref{ddvz0}) and (\ref{dddvz0}),
we have to enforce the usual conditions for the Green function 
at the singularity position $z=z_0$, namely the velocity, its first and
its second derivative are continuous at $z=z_0$. The only discontinuity
comes from the third derivative
\begin{equation}
\frac{\partial^3 \widetilde{v}_z}{\partial z^3} \Big\vert_{z_0^+} -
\frac{\partial^3 \widetilde{v}_z}{\partial z^3} \Big\vert_{z_0^-}  =
\frac{q^2 \widetilde{F}_z}{ \eta_1}
\end{equation}
The algebra involved to evaluate the height integration constants is rather 
lenghty but presents no difficutly. We simply give the resulting
velocity field
\begin{widetext}
\begin{equation}
\widetilde{v}_z (\mathbf{q},z,\omega) = \frac{\widetilde{F}_z}{4 \eta_1 q}
\left[ (1+q \vert z-z_0 \vert)  e^{-q\vert z-z_0\vert }  
 - \left( \frac{1-\lambda }{1+\lambda } \right) \left(1+q(z+z_0)+2q^2zz_0\right)
e^{-q(z+z_0)} \right]    
-\omega_q \widetilde{h}(\mathbf{q},\omega) (1+qz) e^{-qz} \ ,
\label{veln}
\end{equation}
\end{widetext}
for $z \geq 0$,  and
\begin{align}
\widetilde{v}_z (\mathbf{q},z,\omega) =  \frac{\widetilde{F}_z}{4 \eta_2 q} & 
 \left( \frac{2}{1+\lambda } \right)  \left(1+q(z_0-z) \right) e^{q( z-z_0) }  \nonumber \\
&-\omega_q \widetilde{h}(\mathbf{q},\omega) (1-qz ) e^{qz }    \ ,
\label{velninf}
\end{align}
for $z \leq 0$. 
The velocity field still depends on the deformation of the interface,
which itself is a function of the velocity through the closure relation~(\ref{closure}).
Evaluating the velocity~(\ref{veln}) or~(\ref{velninf})
at height $z=0$ and comparing with~(\ref{closure}) then leads to 
\begin{equation}
\widetilde{h}(\mathbf{q},\omega) = \frac{1}{\omega_q + i \omega} 
(1+qz_0) e^{-qz_0}\frac{ \widetilde{F}_z }{4  \overline{\eta} q}    \ .
\label{fdnn}
\end{equation}
Bringing Eq.~(\ref{veln}) and~(\ref{velninf}) together with~(\ref{fdnn}), 
we finally obtain the normal-normal component of
the Green function
\begin{widetext}
\begin{align}
\widetilde{\mathcal{G}}_{zz} (\mathbf{q},z,z_0,\omega) =  \frac{1}{4 \eta_1 q}  &
\left[ (1+q \vert z-z_0 \vert)  e^{-q\vert z-z_0\vert } 
- \left(1+q(z+z_0)+\frac{2q^2zz_0}{1+\lambda}\right)
e^{-q(z+z_0)} \right]   \nonumber \\
&  + \frac{1}{4 \overline{\eta} q} \frac{\omega}{\omega - i \omega_q}
(1+qz)(1+qz_0)e^{-q(z+z_0)}   \ ,
\label{gzzsup}
\end{align}
for $z \geq 0$, and 
\begin{equation}
\widetilde{\mathcal{G}}_{zz} (\mathbf{q},z,z_0,\omega) =  \frac{1}{4 \eta_2 q} 
\left( \frac{2 }{1+\lambda}\right) q^2zz_0 
e^{q( z-z_0) }   
+ \frac{1}{4 \overline{\eta} q} \frac{\omega}{\omega - i \omega_q}
(1 - q z)(1+qz_0)e^{q(z-z_0)}  \ ,
\label{gzzinf}
\end{equation}
for $z\leq 0$. 
\end{widetext}

\subsection{Normal-longitudinal component}

In order to get the component $\widetilde{\mathcal{G}}_{nl}$
of the Oseen tensor, we perform the same analysis
expect that we now keep $\widetilde{F}_l \neq 0$, whereas
we set $\widetilde{F}_z=0$.
This time, the discontinuity imposed by $\delta'$
in Eq.~(\ref{eqtextvz}) has an incidence on 
the \textit{second derivative} of the velocity at $z=z_0$
\begin{equation}
\frac{\partial^2 \widetilde{v}_z}{\partial z^2} \Big\vert_{z_0^+} -
\frac{\partial^2 \widetilde{v}_z}{\partial z^2} \Big\vert_{z_0^-} =
\frac{i q \widetilde{F}_l}{ \eta_1} \ ,
\end{equation}
the velocity, its first and its third derivative being continuous.
The algebra being quite similar to that of the previous section,
we shall skip the details. 
Once again, the velocity field depends on the deformation of the interface.
Intersetingly, a point force exerted \textit{parallel} to the surface is responsible
for a \textit{normal} displacement of the fluid-fluid interface.
Evaluating the velocity at height $z=0$ leads to the result 
\begin{equation}
\widetilde{h}(\mathbf{q},\omega) = \frac{1}{\omega - i \omega_q} 
qz_0 e^{-qz_0}\frac{\widetilde{F}_l}{4 \overline{\eta} q}    \ .
\end{equation}
Bringing 
everything together, we find the normal-longitudinal component of
the Green function
\begin{widetext}
\begin{align}
\widetilde{\mathcal{G}}_{zl} (\mathbf{q},z,z_0,\omega) =  \frac{i}{4 \eta_1 q} &
\left[ q\left( z_0-z \right)  e^{-q\vert z-z_0\vert } 
+ \left(     \frac{1 -\lambda}{1+ \lambda} qz -qz_0-  \frac{2q^2zz_0}{1+\lambda}\right)
e^{-q(z+z_0)} \right]   \nonumber \\
& + \frac{i}{4 \overline{\eta} q} \frac{\omega}{\omega - i \omega_q}
(1+qz)qz_0 e^{-q(z+z_0)}  \ ,
\end{align}
for $z \geq 0$, and 
\begin{equation}
\widetilde{\mathcal{G}}_{zl} (\mathbf{q},z,z_0,\omega) =  -\frac{i}{4 \eta_2 q}
\left( \frac{2}{1+\lambda} \right)
qz(1- qz_0) e^{q( z-z_0) } 
+ \frac{i}{4 \overline{\eta} q} \frac{\omega}{\omega - i \omega_q}
(1 - q z)qz_0e^{q(z-z_0)} \ ,
\end{equation}
for $z \leq 0$. 
\end{widetext}

\section{Longitudinal component of the velocity}
\label{lj}

To obtain the longitudinal component of the velocity,
there is actually no need to solve the corresponding 
differential equation~(\ref{stokesl}).
Indeed, from the incompressibility condition~(\ref{incompressibility}),
$\widetilde{v}_l$ is related to $\widetilde{v}_z$ thanks to
$\widetilde{v}_l = (i/q) 
\partial \widetilde{v}_z / \partial z$.
From the definition of the Oseen tensor
$\widetilde{v}_l=\widetilde{\mathcal{G}}_{ll}\widetilde{F}_l+
\widetilde{\mathcal{G}}_{lz}\widetilde{F}_z$ (since, of course, 
$\widetilde{\mathcal{G}}_{lt}=0$), it is straightforward to get
\begin{widetext}
\begin{align}
\widetilde{\mathcal{G}}_{ll} (\mathbf{q},z,z_0,\omega)
 =  \frac{1}{4 \eta_1 q} &
\left[ \left( 1- q\vert z-z_0 \vert \right)  e^{-q\vert z-z_0\vert } 
- \left(     \frac{1 -\lambda}{1+ \lambda} \right) \left(1-q(z +z_0)+
  \frac{2q^2zz_0}{1-\lambda}\right)
e^{-q(z+z_0)} \right]   \nonumber \\
& + \frac{1}{4 \overline{\eta} q} \frac{\omega}{\omega - i \omega_q}
q^2 zz_0 e^{-q(z+z_0)}  \ ,
\end{align}
for $z \geq 0$, and 
\begin{equation}
\widetilde{\mathcal{G}}_{ll}  (\mathbf{q},z,z_0,\omega)
 =  \frac{1}{4 \eta_2 q} \left(\frac{2}{1+\lambda}\right)(1+qz)
 (1-qz_0) e^{q(z-z_0) } 
+ \frac{1}{4 \overline{\eta} q} \frac{\omega}{\omega - i \omega_q}
q^2 zz_0 e^{q(z-z_0)}  \ ,
\end{equation}
for $z \leq 0$. 
\end{widetext}
Regarding the longitudinal-normal component, 
$\widetilde{\mathcal{G}}_{lz}$, 
no additional algebra is required since it  
can directly be deduced from $\widetilde{\mathcal{G}}_{zl}$
using the symmetry relation of the Green function.


\begin{thebibliography}{99}
\bibitem{remark}
As a matter of fact, the integral can be expressed
in terms of Sine Integrale and Cosine Integral functions, though 
the final result is quite complicated and not really more
transparent than Eq.~(\ref{formula}).
\end{thebibliography}
\end{document}